\documentclass{revtex4-1}

\usepackage{pseudocode}
\usepackage{graphicx}
\usepackage{wasysym}
\usepackage{tikz}

\begin{document}

\title{Exploration of the local solar neighborhood \\ I: Fixed number of probes}

\author{Daniel Cartin}
\email{cartin@naps.edu}
\affiliation{Naval Academy Preparatory School, 440 Meyerkord Ave, Perry Hall, Newport, RI 02841-1519}

\begin{abstract}

Previous work in studying interstellar exploration by one or several probes has focused primarily either on engineering models for a spacecraft targeting a single star system, or large-scale simulations to ascertain the time required for a civilization to completely explore the Milky Way galaxy. In this paper, a simulated annealing algorithm is used to numerically model the exploration of the local interstellar neighborhood (i.e. on the order of ten parsecs of the Solar System) by a fixed number of probes launched from the Solar System; these simulations use the observed masses, positions and spectral classes of targeted stars. Each probe visits a pre-determined list of target systems, maintains a constant cruise speed, and only changes direction from gravitational deflection at each target. From these simulations, it is examined how varying design choices -- differing the maximum cruise speed, number of probes launched, number of target stars to be explored, and probability of avoiding catastrophic system failure per parsec -- change the completion time of the exploration program and the expected number of stars successfully visited. In addition, it is shown that improving this success probability per parsec has diminishing returns beyond a certain point. Future improvements to the model and possible implications are discussed.

\end{abstract}

\maketitle

\section{Introduction}
\label{introduction}

In recent decades, research devoted to interstellar travel and exploration has focused principally in two areas. The first is predominantly aimed toward the design of spacecraft capable of going to nearby stars with current or near-future technology~\cite{large-craft, starwisp}. This has resulted in a creative outpouring of possible methods to accelerate and decelerate these craft, as well as means to communicate scientific data back to the Solar System. On the other hand, there have been a number of theoretical undertakings which examine the time required to engage in an exploration of galactic scale~\cite{bjork, Cot-Mor09, For-Pap-Kit12}. The rationale behind this research, at least in part, is an attempt to better understand the Fermi paradox of SETI. Surveys of this type allow one to estimate the time to completely explore the galaxy, giving an upper bound on the frequency of visits by other civilizations to the Solar System. In particular, Bj\o rk~\cite{bjork} assumes that a single probe is initially sent out, which launches a small (four or eight) set of sub-probes to explore a given sector of the galaxy, with 40,000 nearby stars targeted. Once this exploration is complete, the sub-probes return to their originating probe, which moves to a different stellar neighborhood to start again. Each sub-probe chooses its next target by picking the nearest star not already visited. This work is built upon by Cotta and Morales~\cite{Cot-Mor09}, who use the same basic exploration model, but use various optimization techniques to reduce the total travel times of the sub-probes. After such a trial optimization, any resulting tour reducing the path time is kept, and the procedure is repeated. These heuristics give roughly a 10\% decrease in the time to explore a given section of the galaxy. Finally, Forgan, Papadogiannakis and Kitching~\cite{For-Pap-Kit12} consider the use of a single probe, again using the choice of the nearest neighboring star as the next target. However, unlike the other works, Forgan et al. use gravitational slingshots from the relative motion of probe and target star to progressively increase the speed of the probe in two of their simulation scenarios, resulting in large increases in that speed.

In this paper, the focus is on the middle ground between these two extremes. It is assumed in particular that the technological means and ability has been developed to launch exploration probes to star systems within a few parsecs, but not far enough into the future that large portions of the Milky Way Galaxy -- or even our local stellar neighborhood -- have been explored. Note that Moir and Barr~\cite{Moi-Bar05} develop results similar to those in this paper, although they are examining the possibility of a spacecraft using a cyclic trajectory to travel to a few nearby stars, then return to the Solar System. The possibility of self-replicating probes is not examined here, so the number of probes is fixed by the quantity launched from the Solar System. To start with, this reduces the number of parameters to consider, namely the number of daughter probes produced at each target system by the original arriving spacecraft. This is not to make any argument for or against self-replicating probes, merely to suggest that it is certainly easier to make a probe capable of traveling to another star system, than it is to build a probe that can do this {\it and} construct perfect replicas of itself upon reaching that system. Thus, it is at least feasible to say the first exploration program of the type considered here will be of a fixed number of probes. An alternate program of self-replicating probes will be saved for a companion paper to the present one.

Our exploration model makes the following additional assumptions:
\begin{enumerate}
	
	\item Probes change direction only using gravitational deflections at each target system. In multiple star systems, the probe may use any of the stars present to optimize the resulting course correction, although only one is used for this deflection (this counts as a visit to the entire multiple star system).

	\item Each probe is accelerated up to a fixed cruise speed while leaving the Solar System, and maintains this speed throughout it assigned path. For simplicity, it is assumed that the positions of the target systems are known in advance, and their proper motions are neglected. Thus, increases in probe speed due to gravitational assists are not considered, as is done by Forgan et al.~\cite{For-Pap-Kit12}.
	
	\item At the time of launch, a given probe will be assigned one or more systems to visit, and will travel through these systems without stopping.

\end{enumerate}
Also, unless another source is cited, the stellar data used -- specifically, the position coordinates, spectral types and masses -- is from the Research Consortium on Nearby Stars database~\cite{RECONS}, and only the sixty nearest systems to the Solar System at most are considered.

Our current lack of experience in launching and utilizing interstellar probes means that the engineering model chosen is up to the individual. There are two typical categories for such a model:

\begin{itemize}

	\item {\it Large spacecraft:} Examples of this include Project Daedalus, its successor Icarus, and the Longshot effort~\cite{large-craft}. These are likely to be more robust under the ravages of radiation damage and collisions with interstellar matter (such as planetesimals ejected from orbits around nearby stars), but are also more expensive in time and resources to construct. On the other hand, they are self-contained, and do not require additional infrastructure beyond construction facilities.
	
	\item {\it Small spacecraft:} This class of interstellar probes includes the ``starwisp" probe, launched and powered using light beams from devices remaining in the Solar System~\cite{starwisp}. Because the power source does not move with the craft, these probes have very low mass, and thus can be produced in great numbers. However, it is unclear how such a lightweight construction will fare over interstellar travel of a few decades, since there would be little to no allowance for shielding against impact damage.

\end{itemize}
Because of these considerations, a conservative choice is made  -- the assumption is that all probes launched are ``large", although some of the results presented below may apply to a choice of ``small" probe size.

One consequence of this choice is the need to use the probes as efficiently as possible, in order not to waste resources. If the construction of a single probe requires a good deal of resources, then the launch of a multi-probe exploration program is likely to be delayed, due to the need to assemble additional resources to construct all probes. Suppose, for example, that a single prototype probe is launched to a nearby star, and the design successfully reaches its destination and returns quality data about the target system. How long would it require to build additional probes to examine other systems? If it is supposed the resource cost per probe as a fraction of either national or world production is kept fixed, then the waiting time may be significant -- with the cost at a fixed percentage, an additional $N$ probes would not be completed until economic production has grown by the same factor of $N$ (see Millis~\cite{Mil-11} and references therein). Table \ref{resource-cost} shows the waiting period necessary to produce a certain additional number of probes, given an annual growth rate in the appropriate measure -- such as the size of a national or world economy, energy production, or other metrics. These values are obtained by a compounding-type equation, namely $R_{probe} = R_1 (1 + c)^t$, where $c$ is the annual growth rate in applicable resources, $t$ the number of years, $R_{probe}$ the resources to construct the desired number of probes $N_{probe}$, and $R_1$ the resources to complete the initial probe. It is assumed that $R_{probe} / R_1 \propto N_{probe}$, i.e. there are no economies of scale when building multiple probes. Note that since this refers to the time to assemble resources for building all probes, it is possible to either build these craft either sequentially (so that the launch times are staggered) or simultaneously. In either case, since the total construction times listed in Table \ref{resource-cost} are small compared to the total travel times between stars, these construction times are not included in the mission completion times given below in Section \ref{results}.

\begin{table}
\begin{tabular}{|c|c|c|c|}
\hline
					& \multicolumn{3}{|c|}{Annual growth rate} \\
$N_{probe}$				&	1 \%		& 3 \%	& 5\%	\\
\hline
\hline
3	& 110 yr			& 37.2 yr		& 22.5 yr		\\
\hline
5	& 162 yr			& 54.4 yr		& 33.0 yr		\\
\hline
10	& 231 yr			& 77.9 yr		& 47.2 yr		\\
\hline
20	& 301 yr			& 101 yr		& 61.4 yr		\\
\hline
\end{tabular}
\caption{\label{resource-cost}Number of years after the launch of a single ``large" prototype probe to build an additional $N_{probe}$ probes, assuming a range of annual growth rates in a relevant quantity, e.g. energy production or financial means.}
\end{table}

\section{Gravitational deflections}
\label{model}

\begin{figure}
\begin{tikzpicture}[>=stealth]


	\draw [very thick] plot[variable=\t, samples=1000, domain = -37:37] ({tan(2 * \t)},{-2.5 + 1.5 * sec(2 * \t)});
	
	
	\filldraw[gray] (0, 0) circle (3pt) node [left = 1pt, black] {$M_s$};
	
	
	\draw[<->] (0, 0) -- (0, -1) node[above = 1.5em, left = 1pt] {$r_p$};
	\draw[<->] (0, -1) -- (0, -2.5) node [above = 3em, left = 1pt] {$a$};
	
	
	\draw[dashed] (0, -2.5) -- (5.5 / 1.5, 3);
	\draw[dashed] (-5.5 / 1.5, 3) -- (0, -2.5);
	\draw[dashed] (0, -2.5) -- (1 / 1.5, -3.5);
	
	
	\draw[dotted, ->] (0.277, -2.5 - 1.5 * 0.277) arc (-56:56:0.5) node [below = 1em, right = 0.75em] {$\Delta \theta$};
	
	
	\draw[->] (-3, 3) -- (-3 + 1 / 1.5, 2);
	\draw[->] (3 - 1 / 1.5, 2) -- (3, 3);
	
	
	\draw[dashed] (0, 0) -- (0.5, 0.75);
	\draw[dotted, ->] (0, -0.5) arc (-90:56:0.5) node [below = 1em, right = 0.75em] {$\theta_\infty$};
\end{tikzpicture}
\caption{\label{grav-defl}Plot of the hyperbolic orbit of a probe undergoing a gravitational deflection due to the mass $M_s$. The perihelion distance $r_p$ and semi-major axis $a$ of the orbit are shown, as well as the angular deflection of the probe; the angle $\theta = 0$ is fixed by the perihelion point of the spacecraft. The probe enters the target system from the upper left, changes direction by $\Delta \theta$, then leaves the system out the upper right. From the diagram, one can see the smaller angle between the asymptotes of the hyperbola can be found in two ways -- namely, $2 \pi - 2 \theta_\infty$, and $\pi - \Delta \theta$. Equating these together and solving for the angular deflection gives $\Delta \theta = 2 \theta_\infty - \pi$.}
\end{figure}
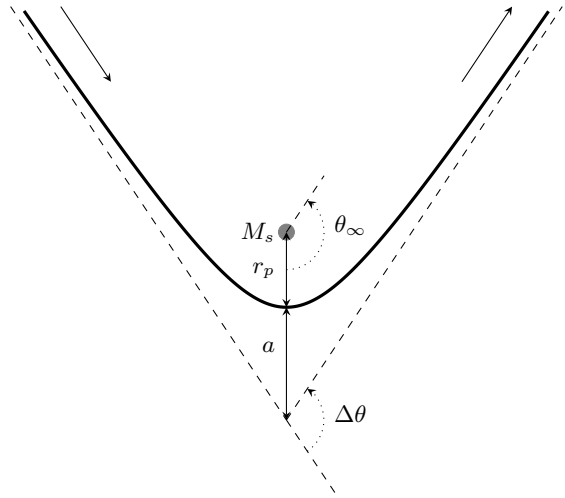

Now the issue of gravitational deflections is taken up, by considering hyperbolic orbits around a star~\cite{Moi-Bar05}. This allows us to find the relationship between the deflection angle between the incoming and outgoing legs of this orbit, and the asymptotic speed of the probe traveling this path. The following derivations are based on the well-known theory of central force orbits (see, e.g.~\cite{Mar-Tho88}). For any conic section, the equation relating the distance $r$ from the focus at a given angle $\theta$ is given by
\begin{equation}
\label{r-theta}
	\frac{\alpha}{r} = 1 + \epsilon \cos \theta
\end{equation}
Here, $2 \alpha$ is the latus rectum of the orbit and $\epsilon$ is the orbit eccentricity. Note that $\theta = 0$ is where the orbit has its closest approach to the focus, so that the perihelion distance $r_p$ is given by
\begin{equation}
\label{rp-eqn}
	r_p = \frac{\alpha}{1 + \epsilon}
\end{equation}
For gravitational deflections, the probe is on a hyperbolic orbit through the target system, so $\epsilon > 1$. As the spacecraft leaves this system, it approaches $r \to \infty$ as the angle reaches $\theta \to \theta_\infty$, the angular position of the probe on the outgoing leg. The value of $\theta_\infty$ is found from the conic section equation (\ref{r-theta}) by
\begin{equation}
\label{max-angle}
	\theta_\infty = \cos^{-1} \biggl(- \frac{1}{\epsilon} \biggr)
\end{equation}
with the total deflection $\Delta \theta = 2 \theta_\infty - \pi$ (see Figure \ref{grav-defl}). For our purposes, it is more useful to have a relation between the maximum cruise speed possible for a given deflection angle $\Delta \theta$. To do this, two equations for the semi-major axis of a hyperbolic orbit are appropriate, namely
\[
	a = \frac{r_p}{\epsilon - 1} \simeq \frac{GM_s}{v^2 _\infty}
\]
where $M_s$ is the mass of the object providing the gravitational deflection (neglecting the probe's mass), and $v_\infty$ is the probe's speed as $r \to \infty$. Combining these relations, along with the above equations (\ref{r-theta}) through (\ref{max-angle}), to eliminate the eccentricity from the relations, gives the result
\begin{equation}
\label{v-inf-eqn}
	v_\infty = \sqrt{\biggl( \frac{GM_s}{r_p}  \biggr) \biggl[ \csc\biggl( \frac{\Delta \theta}{2} \biggr) - 1 \biggr]}
\end{equation}
Thus, for a desired deflection $\Delta \theta$ to the next target star, the maximum possible speed the probe can have depends on the geometric factor $\Delta \theta$, and a constant $\sqrt{GM_s / r_p}$ related to the star itself. Any probe cruise speed $v \le v_\infty$ will allow the spacecraft to execute a corresponding angular deflection of $\Delta \theta$. This will be a factor in our results -- it may be technologically feasible to accelerate a probe for a wide range of cruise speeds, but the probe will be limited to the maximum speed allowed by the path mapped out for it between target stars, and so the craft's cruise speed will be bounded above by the smallest value of $v_\infty$ along its trajectory.

Turning next to how the constant $\sqrt{GM_s/r_p}$ scales with different stellar properties, writing it in terms of the Sun's mass $M_{\astrosun}$ and radius $r_{\astrosun}$ gives
\begin{equation}
\label{const-eqn}
	\sqrt{\frac{GM_s}{r_p} } = \biggl[ 4.368 \times 10^5 \biggl( \frac{M_s}{M_{\astrosun} } \biggr)^{1/2} \biggr( \frac{r_p}{r_{\astrosun} } \biggr)^{-1/2} \biggr] \textrm{m/s} = [0.01457\ c] \biggl( \frac{M_s}{M_{\astrosun} } \biggr)^{1/2} \biggr( \frac{r_p}{r_{\astrosun} } \biggr)^{-1/2}
\end{equation}
The perihelion distance is fixed by the need to limit the amount of heat flux incident on the stellar probe due to stellar radiation, which is related to the (bolometric) luminosity $L_s$ of the star and the desired maximum heat flux $H_{max}$ falling on the spacecraft. As discussed in Moir and Barr~\cite{Moi-Bar05}, bounding the flux incident on the probe at small distances to the star helps to limit the total heat load on the craft. Perihelion distance and maximum heat flux are related by
\begin{equation}
\label{min-dist-eqn}
	r_p = \sqrt{ \frac{L_s}{4 \pi H_{max} } }
\end{equation}
To show which types of main-sequence stars provide the most useful gravitational deflections, the empirical relation~\cite{Lang}
\begin{equation}
\label{mass-lum-eqn}
	\biggl( \frac{L_s}{L_{\astrosun} } \biggr) = \biggl( \frac{M_s}{M_{\astrosun} }\biggr)^\alpha
\end{equation}
is used between stellar luminosity $L_s$ and mass $M_s$ of a main-sequence star, in terms of the comparable properties for the Sun; the exponent $\alpha$ is determined from astronomical data to be in the range $\alpha \simeq 4$. Using the value of $H_{max} = $ 7 MW/m$^2$ of Moir and Barr, equations (\ref{const-eqn}) through (\ref{mass-lum-eqn}) give
\begin{equation}
\label{const-eqn-2}
	\sqrt{\frac{GM_s}{r_p} } = \biggl( \frac{4 \pi G^2 H_{max} M^\alpha _{\astrosun} }{L_{\astrosun} } \biggr)^{1/4} M_s ^{(2 - \alpha)/4} = [8.525 \times 10^{-4} \ c] \biggl( \frac{M_s}{M_{\astrosun} } \biggr)^{(2 - \alpha)/4}
\end{equation}
Since $\alpha > 2$ for main-sequence stars, then as mass decreases, the maximum cruise speed $v_\infty$ increases. Note that -- although we use the fixed value $H_{max} = 7$ MW/m$^2$ for the maximum heat flux throughout this work -- improvements in the radiation shielding of the probes during close approaches may increase the relative size of $v_\infty$, as $\sqrt{GM_s / r_p} \propto H _{max} ^{1/4}$.

Less massive main-sequence stars produce less energy from nuclear fusion for a given mass than the larger variety. Thus, redder stars serve to provide the greatest angular change in a probe trajectory from solely gravitational attraction, while bluer stars provide the least. White dwarfs are even better, since their radiant energy does not arise from nuclear fusion and they have high mass densities, as their matter is supported only by electron degeneracy. Brown dwarfs are another favorable object class -- although they are not as dense as white dwarfs, they do not emit much radiation due to their lack of nuclear fusion. Specific values of the constant $\sqrt{GM_s / r_p}$ are given for a variety of nearby objects in Table \ref{star-constant-list}. Note that the use of only gravitational assists means that some systems may be targeted solely for their ability to change a probe's course. An example of this is van Maanen's Star, a solitary white dwarf which has a relatively high constant; thus, although it may be of inherent astrophysical interest, it also serves as a useful waypoint to create rather large angular deflections. In addition, when a probe is incident on a multiple star system, the usual best choice of star to use for a gravitational deflection will be the reddest of the stars present, if not a brown or white dwarf. For example, as shown in Table \ref{star-constant-list}, a much greater change results from a close pass near Sirius B, rather than the brighter Sirius A. However, this aspect can be problematic in other situations, e.g. the $\alpha$ Centauri star system, where the best choice for such a deflection -- Proxima Centauri -- lies at a relatively large distance ($\sim 0.237$ light-years) from the other two members of this triple star system. Since this is the sole example in the selection of target stars considered here, for simplicity, we count a probe visiting Proxima Centauri as one to the entire $\alpha$ Centauri system.

\begin{table}
\begin{tabular}{|c|c|c|c|}
\hline
Star name						& Stellar classification	& $\sqrt{GM_s / r_p}$ (fraction of $c$)\\
\hline
\hline
Sirius A~\cite{sirius2}					& A1V			& $5.32 \times 10^{-4}$		\\
\hline
Fomalhaut~\cite{fomal}				& A3V			& $5.77 \times 10^{-4}$		\\
\hline
Procyon A	~\cite{solar}				& F5IV-V		& $6.30 \times 10^{-4}$		\\
\hline
$\alpha$ Centauri A~\cite{solar}			& G2V			& $8.00 \times 10^{-4}$		\\
\hline
$\alpha$ Centauri B~\cite{solar}			& K1V			& $9.81 \times 10^{-4}$		\\
\hline
Barnard's Star~\cite{Barnard}			& M4V			& $1.38 \times 10^{-3}$		\\
\hline
Proxima Centauri~\cite{proxima}			& M5.5V		& $1.44 \times 10^{-3}$		\\
\hline
Sirius B~\cite{sirius1, sirius2}				& DA2			& $2.17 \times 10^{-3}$		\\
\hline
$\epsilon$ Indi Ba~\cite{eps-Indi}			& T1			& $3.01 - 3.32 \times 10^{-3}$	\\
\hline
$\epsilon$ Indi Bb~\cite{eps-Indi}			& T6			& $3.62 - 4.09 \times 10^{-3}$	\\
\hline
Procyon B~\cite{procyon}				& DQZ			& $4.38 \times 10^{-3}$		\\
\hline
van Maanen's Star~\cite{maanen1, maanen2}	& DZ7			& $5.62 \times 10^{-3}$		\\
\hline
\end{tabular}
\caption{\label{star-constant-list}Values of the constant $\sqrt{GM_s / r_p}$ for a variety of nearby stars, brown dwarfs and white dwarfs, where $M_s$ is the mass of the object, and $r_p$ is the perihelion distance for the interstellar probe, determined by using equations (\ref{const-eqn}) and (\ref{min-dist-eqn}) with a maximum desired heat flux of 7.00 MW/m$^2$. Mas and luminosity data for each star taken from the references listed; for Procyon B, Proxima Centauri, Sirius B, and van Maanen's Star, the luminosity was derived from the Stefan-Boltzmann law, using the radii and effective temperatures provided in the references. The range in the values for the brown dwarfs $\epsilon$ Indi Ba and Bb represent the observational uncertainty in the mass of those objects.}
\end{table}

\section{Results}
\label{results}

An efficient exploration program is now sought, using a fixed number of probes launched from the Solar System at the same time, by simulating the results of different design choices. By ``efficiency", the amount of time to complete the program is minimized in some sense. For example, with $t_a$ the time for probe $a$ to complete its portion of the mission, our efficiency metric could be the largest of the values $t_a$, or the sum of the completion times $\sum_a t_a$; for the sake of definitiveness, the simulations presented below will minimize the Euclidean norm of all $t_a$, e.g. an objective function
\begin{equation}
\label{obj-func}
	C = \min \sum_a t_a ^2
\end{equation}
This measure is aimed at producing probe travel times $t_a$ that are roughly comparable in size.
The possible variables of the program include how many probes $N_{probe}$ are launched as part of the exploration program, the cruise speed for each of the probes (which may not be the same for all spacecraft), the probability $p$ of each probe successfully traveling a given parsec, and the total number of target systems $N_{target}$. These variables are now discussed in turn. Building a substantial number of probes would require significant economic resources; Thus, an exploration program of large spacecraft would seek to minimize $N_{probe}$ as much as possible, while trying simultaneously increasing the number of targets $N_{target}$ successfully reached in the shortest feasible time. The probe cruise speed is bounded by the path chosen for each path; a separate consideration is whether the spacecraft can be actually accelerated up to a given velocity. Thus, the specific parameter used in the simulations is the maximum possible cruise speed $v_p$ the probes can use. Specifically, this means the probes can use any path, where for all angular deflections $\Delta \theta$ the maximum speed $v_\infty$ -- given by the relation (\ref{v-inf-eqn}) between them -- does not exceed $v_p$. In other words, the parameter $v_p$ serves to bound the search space of possible paths. However the largest significance of this is on the travel time of the probe; although a probe with a large cruise speed will not be physically able to make a large angular change in its path, it can conceivably recover this by going faster along its trajectory.

Finally, the chance of a catastrophic failure of a probe is considered, leading to its inability to complete its mission. For the following, it is assumed that the probability $p$ of a probe completely failing for a given parsec of its path is independent of it failing along any other parsec distance. This is justified by the following. Since the probe has a constant cruise speed, the time of flight is directly proportional to the distance traveled. This means that intrinsically time-related failures -- such as critical systems onboard the probe -- will have a path distance relation just as much as physical phenomena associated with the trajectory itself -- e.g. the probability for the probe shielding to fail due to a high-energy impact of cosmic dust. It is assumed that the time-dependent and distance-dependent failure modes have similar probability distributions, so that the net effect is that the success probability $p$ for a craft moving at constant speed is a probability distribution per parsec.

\begin{figure}
	\includegraphics[width=0.6\textwidth]{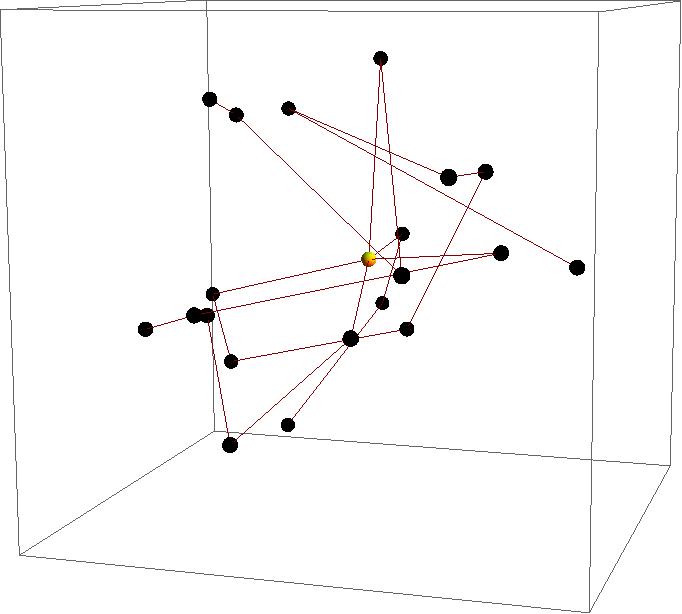}
	\caption{\label{example-sim}An example of a simulated exploration program, for $N_{target} = 20$ and $N_{probe} = 5$, so that all star systems within 3.68 pc of the Solar System are surveyed. Lines represent the path of a probe from one system to another; the Solar System is the central node with five paths emerging from it. The first probe completes its mission after 1289 years, the last after 3931 years.}
\end{figure}

Varying the parameter $p$ will change the expected number of target systems successfully reached, and the time it takes to do so -- either the time per probe, or the time to complete the entire exploration program. These results are below are arrived at by using a simulated annealing algorithm~\cite{sim-anneal} to choose the best path for each probe. Although this does not guarantee the optimal routes for a multi-probe exploration program, it does provide a solution that is relatively good in a short amount of computational time. Further details about the algorithm, and its effectiveness compared to the exact optimal solutions and other heuristic approaches, are in the Appendix. An example simulation is demonstrated in Figure \ref{example-sim}.

\begin{figure}
	\includegraphics[width=0.6\textwidth]{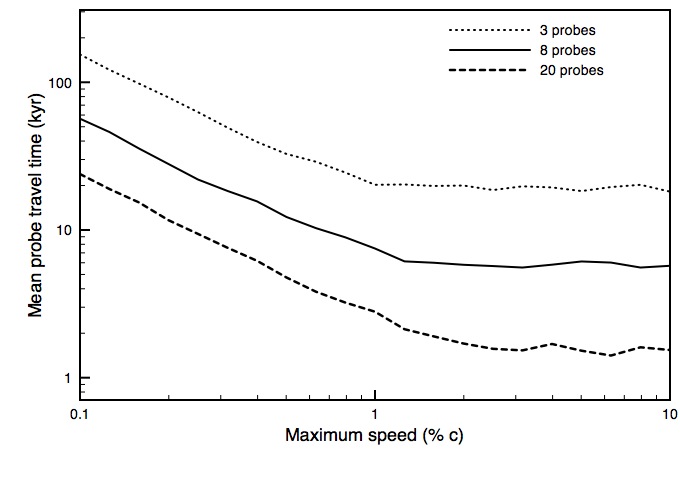}
	\caption{\label{max-speed-graph}Mean travel time for each probe in a collection of probes, as a function of the cruise speed $v_p$ for each probe. Three possible choices of the number of probes launched from the Solar System is given, for a total of $N_{target} = 60$ systems are visited by the collection.}
\end{figure}

The first property of the exploration probes considered is the upper bound $v_p$ on their cruise speed. For the large probes considered here, this speed will depend on the method of launch from the Solar System -- is the method of propulsion onboard, or are other methods, e.g. beamed propulsion, used? Without getting into particulars, the effect of varying the given cruise speed has on the mean time for each probe to complete its exploration program is examined. As seen in Table \ref{star-constant-list}, the values of $\sqrt{GM_s / r_p}$ for various stellar types sets the scale of the probe cruise speeds inside the range $10^{-2} c - 10^{-4} c.$ A probe exceeding this range may have shorter travel times, but also a more limited selection of stars to aim towards after leaving a target system, since the required course deflection may not be possible at its cruise speed. In other words, a desired exploration path for a given probe is likely to fix a maximum possible cruise speed to be on the order of $10^{-2}c$. This varies according to the number of probes $N_{probe}$ -- if more probes are launched for a fixed number of target systems, each probe has to visit fewer systems, leading to less stringent limitations on cruise speed. Thus, in Figure \ref{max-speed-graph}, for $N_{probe} = 3$ and $N_{probe} = 8$, the travel time plateaus around $v_p = 0.01c$, but for $N_{probe} = 20$, the travel time decreases until about $v_p = 0.02c$. For sake of definitiveness, the maximum cruise speed of all probes is set to $v_p = 0.01c$ in the remainder of this paper.

\begin{figure}
	\includegraphics[width=0.6\textwidth]{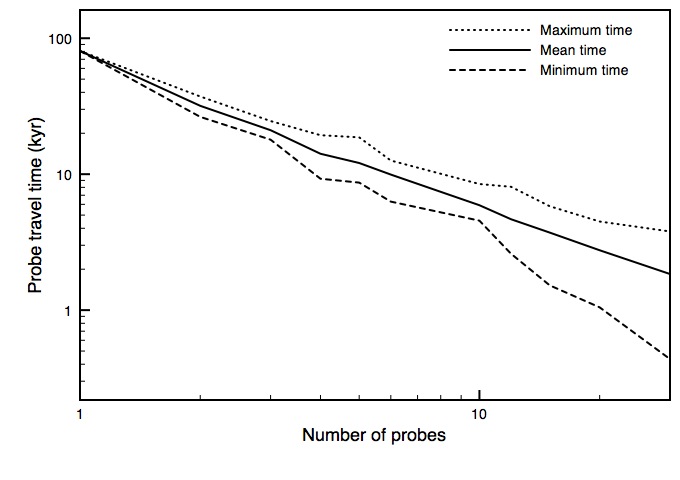}
	\caption{\label{probe-num-graph}Minimum, mean and maximum travel times for probes exploring a total of $N_{target} = 60$ systems, as a function of the total number of probes $N_{probe}$ used in the exploration program.}
\end{figure}

Next, the effect of increasing the total number of probes $N_{probe}$ on the travel times of each probe is studied; for $N_{target} = 60$, the minimum, mean and maximum travel times for the collection of probes are shown in Figure \ref{probe-num-graph}. Recall that the objective function used is the Euclidean norm (\ref{obj-func}) of all probe completion times, so various other measures of the travel times are reported in Figure \ref{probe-num-graph} as a different viewpoint on the solutions arrived at. Somewhat surprisingly, all three of these measures are roughly linear in this log-log plot, meaning that $t \propto p^k$, for some power $k$. Note that for most simulations used to obtain a scaling law of this type, $N_{probe} \ll N_{target}$, since for comparatively large numbers of probes, the simulated annealing algorithms would not assign any target systems to a small percentage of the spacecraft; simulations with these results were discarded. Simulations over a large variety of choices for $v_p, N_{probe}$ and $N_{target}$ were run; specifically, $v_p$ ranged from $10^{-3} - 10^{-2}c$, $N_{target}$ took values of multiples of ten within the range $[10, 60]$, and values of $N_{probe}$ were all multiples of four such that the simulation algorithm gave assigned paths to all probes. From these simulations, it was found that the mean travel time for each probe is given by the scaling law
\begin{equation}
\label{scaling-mean}
	t_{mean} \propto N_{target} ^{0.935 \pm 0.0208} v_p ^{-0.942 \pm 0.0152} N_{probe} ^{-1.04 \pm 0.0213}
\end{equation}
For minimum and maximum times,
\begin{equation}
	t_{min} \propto  N_{target} ^{1.30 \pm 0.0262} v_p ^{-0.921 \pm 0.0192} N_{probe} ^{-1.27 \pm 0.0268}
\end{equation}
and
\begin{equation}
\label{scaling-max}
	t_{max} \propto N_{target} ^{1.01 \pm 0.00974} v_p ^{-0.929 \pm 0.00714} N_{probe} ^{-0.797 \pm 0.00998}
\end{equation}
respectively. All of these relations were found by using a multiple linear regression on the logarithms of the variables; the coefficient of determination $R^2$ for all three equations exceeds 0.934. From these relations, all time scales are seen to be inversely proportional to the maximum speed of the probes; at least roughly, staying within our chosen range $v_p \le 0.01c$, doubling this maximum speed causes the mean time for each probe to complete its mission to be halved. The dependence of the time on the number of target stars and probes has a wider variance. Specifically, doubling $N_{target}$ reduces the minimum time for a probe to complete its mission by 58.5\%, but the maximum time only by 42.4\%. Finally, although it is not included in the equations listed above, the various times $t$ scale with the heat flux $H_{max}$ incident on the probe at perihelion as $t \propto H_{max} ^{1/4}$, as seen by the form (\ref{const-eqn-2}) of the constant $\sqrt{GM_s / r_p}$ and verified by numerical simulations.

\begin{figure}
	\includegraphics[width=0.6\textwidth]{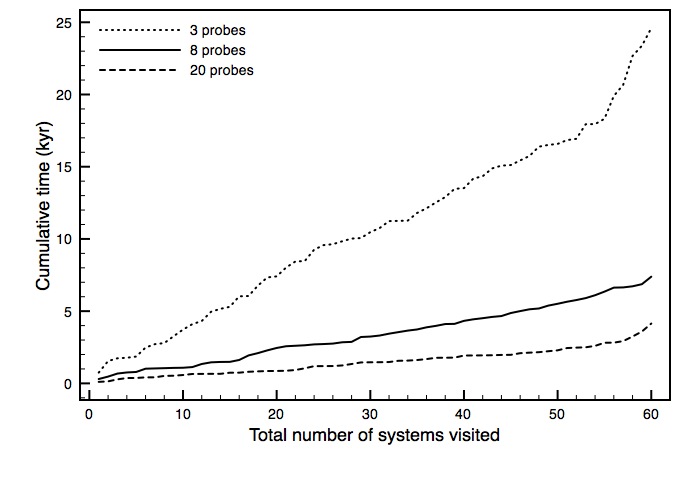}
	\caption{\label{cum-time-graph}Cumulative time to visit a given number of systems, in units of kiloyears, with either $N_{probe} = 3, 8,$ or $20$ probes used to explore a total of $N_{target} = 60$ systems.}
\end{figure}

\begin{figure}
		\includegraphics[width=0.6\textwidth]{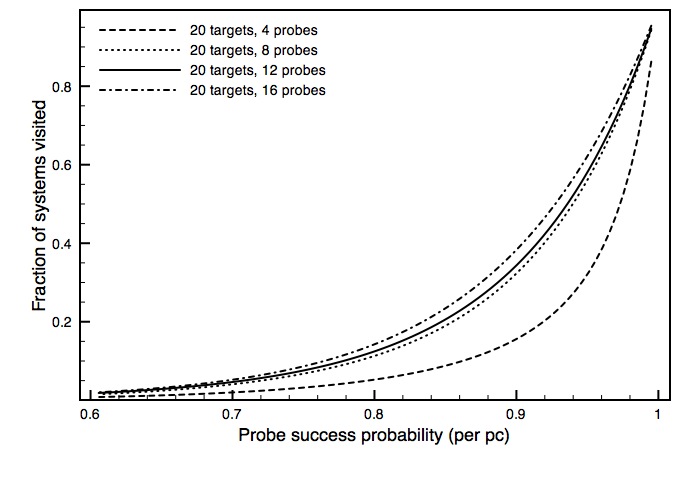}
		\includegraphics[width=0.6\textwidth]{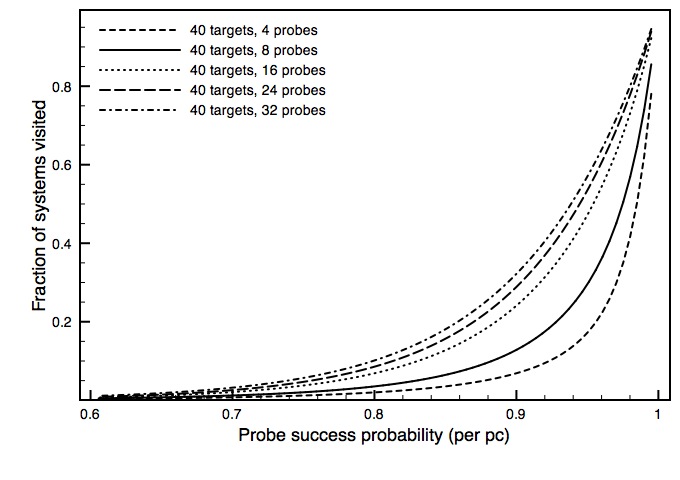}
		\includegraphics[width=0.6\textwidth]{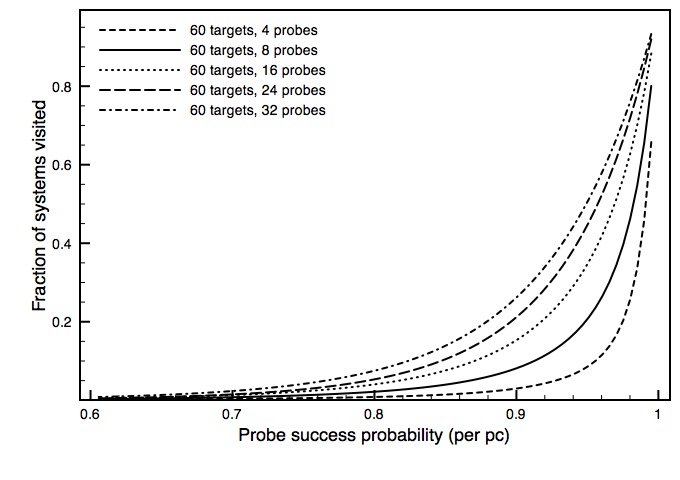}
	\caption{\label{probe-prob-graph}Graphs of the expected number of systems $\langle N \rangle$ visited by a collection of probes launched from the Solar System, as a function of the probability $p$ to successfully cross each parsec without catastrophic failure. Each plot corresponds to a different number of total systems $N_{target}$ to be visited; within the plot, difference choices for the number of probes $N_{probe}$ launched is given.}
\end{figure}

Up until this point, all results presented have assumed that the probes successfully carry out their assigned missions without mishap. Obviously, this is a simplifying assumption, since there would be a non-zero probability of catastrophic failure for each probe. The probability of the probe successfully traveling an additional parsec along its path is denoted as $p$. Therefore, supposing the distance between systems $i$ and $j$ in parsecs is given by $d_{ij}$, the probability for a probe to successfully cross this distance is given by $p^{d_{ij}}$, while the probability of catastrophic failure is $1 - p^{d_{ij}}$. Now the expected number $\langle N \rangle$ of systems visited by the collection of probes is sought. Since all probes are independent, the expected number of systems $\langle N \rangle$ visited by all probes is simply the sum $\sum_a \langle N_a \rangle$ of systems visited by each probe $a$. With $k_a$ systems for probe $a$ to visit, the expected number of systems visited is defined as
\begin{equation}
\label{exp-num}
	\langle N_a \rangle = \sum_{n = 0} ^{k_a} n p_n 
\end{equation}
where $p_n$ is the probability of reaching a maximum of $n$ systems out of the $k_a$ possible, with $\sum_n p_n = 1$. The value $p_n$ is computed using a formal power series in a variable $x$, where the coefficient of the $x^n$ term will indicate the probability of $n$ being the expected number of systems successfully visited. The probability of the probe going from system $i$ to system $j$ thus is represented by the linear function
\[
	(1 - p^{d_{ij}}) + p^{d_{ij}} x
\]
while the outcome of going from $i \to j \to k$ is given by the combination of two such functions, namely
\begin{equation}
\label{two-probe-prob}
	(1 - p^{d_{ij}}) + p^{d_{ij}} x [(1 - p^{d_{jk}}) + p^{d_{jk}} x] = (1 - p^{d_{ij}}) + p^{d_{ij}} (1 - p^{d_{jk}}) x + p^{d_{ij}} p^{d_{jk}} x^2
\end{equation}
Again, each coefficient in the right-hand side of equation (\ref{two-probe-prob}) gives the probability for each value of the expected number of systems visited. In particular, the constant value $(1 - p^{d_{ij}})$ gives the chance that no systems are visited, since it shows the probability that the probe does not successfully travel from $i$ to $j$. The term linear in $x$ gives the probability that one system is reached, since the probe reaches system $j$ -- with chance $p^{d_{ij}}$ -- but not system $k$ -- with chance $1 - p^{d_{jk}}$. Finally, the quadratic term shows the probability of reaching both system $j$ (probability $p^{d_{ij}}$) and then system $k$ (probability $p^{d_{jk}}$). Note that this probability is the same as $p^{d_{ij} + d_{jk}}$, i.e. the parameter $p$ raises to the power of the total distance traveled along the probe's path. Thus, if the distance $d_i$ is defined as the distance along the path of the probe, from the Solar System (node $0$) to system $i$, then the function $f_a (x)$ giving the expected number of systems visited by probe $a$ is
\begin{eqnarray*}
	f_a (x) &=& (1 - p^{d_1}) + p^{d_1} (1 - p^{d_2}) x +  \cdots + p^{d_{k_a - 1}} (1 - p^{d_{k_a}}) x^{k_a - 1} + p^{d_{k_a}} x^{k_a} \\
		&=&  \biggl[ \sum_{n = 0} ^{k_a - 1} p^{d_n} (1 - p^{d_{n + 1}}) x^n \biggr] + p^{d_{k_a}} x^{k_a}
\end{eqnarray*}
where $d_0 = 0$. The last term in the series is different than the others, since it is irrelevant if the probe survives past the last targeted system on its path. In order to arrive back at the expected number $\langle N_a \rangle$ given by (\ref{exp-num}), note that
\[
	\langle N_a \rangle = \frac{df_a}{dx} \biggl |_{x=1}
\]
To find the expected number $\langle N \rangle$ of systems visited by all probes, by the chain rule for derivatives,
\[
	\langle N \rangle = \frac{d}{dx} \biggl[ \prod_a f_a (x) \biggr]_{x=1}
\]
Using this procedure, the variation in the expected number of systems visited for particular choices of $N_{target}$ and $N_{probe}$ are shown in Figure {\ref{probe-prob-graph}}. The probabilities $p$ used in computing these results were chosen as $p = 0.995^k$, for integer values of $k$. Thus there may be differences with the discussion following in the case of $0.995 < p < 1$. However, for all cases calculated with $p = 0.995, \langle N \rangle \sim N_{target}$, so it is likely that the dependence of the expected number of visited systems on $N_{probe}$ is almost nonexistent for such high success probabilities.

One interesting result is how increasing the number $N_{probe}$ of probes launched has limited utility beyond a certain value. To show a particular instance of this, suppose a probe design is chosen where the probability of success per parsec is 0.90; the resulting expected number of systems visited, given a particular number of systems $N_{target}$ targeted, is shown in Table {\ref{fixed-prob}}. In particular, increasing the desired number of systems actually visited requires large numbers of extra probes. For example, for $N_{target} = 20$, one has to double the number of probes from $N_{probe} = 8$ to $N_{probe} = 16$ to gain only one additional visited system. For the choice of $N_{target} = 60$, results improve somewhat -- doubling the number of probes from $N_{probe} = 16$ to $N_{probe} = 32$ results in a 70\% increase in the expected number of systems visited -- but further increases beyond $N_{probe} = 32$ do not give any appreciable gain beyond $\sim 15$ systems. To express this in a different form, suppose $N_{probe} = 60$ systems are targeted, and one wishes to see what success probability $p$ is necessary to ensure probes arrive at a minimum of $\langle N \rangle = 10$ systems; these probabilities are given in Table \ref{fixed-outcome}. Consider the situation where one starts with a projected program of $N_{probe} \ge 28$, but would rather launch fewer probes by expending engineering effort to increase the probe success probability $p$. Table \ref{fixed-outcome} shows that, at first, small changes in $p$ are needed to reduce $N_{probe}$, but that diminished returns occur as $N_{probe}$ decreases below $N_{probe} \le 16$.

\begin{table}
\begin{tabular}{|c|c|c|c|c|c|c|}
\hline
			& \multicolumn{6}{|c|}{$N_{probe}$}				\\
$N_{target}$	& 4		& 8		& 12		& 16		& 24		& 32		\\
\hline
\hline
20			& 3.13	& 6.46	& 6.88	& 7.67	& --		& --		\\
\hline
40			& 2.75	& 5.14	& 8.54	& 9.65	& 11.6	& 12.9	\\
\hline
60			& 1.79	& 4.89	& 5.81	& 9.19	& 12.7	& 15.7	\\
\hline
\end{tabular}
\caption{\label{fixed-prob}Expected number $\langle N \rangle$ of systems visited as a function of the number of systems $N_{target}$ targeted and the number of probes $N_{probe}$ launched, for a fixed success probability per parsec of $p = 0.90$.}
\end{table}

\begin{table}
\begin{tabular}{|c|c|}
\hline
$N_{probe}$	& $p$	\\
\hline
\hline
4			& 0.971	\\
\hline
8			& 0.940	\\
\hline
12			& 0.930	\\
\hline
16			& 0.906	\\
\hline
20			& 0.891	\\
\hline
24			& 0.883	\\
\hline
28			& 0.867	\\
\hline
32			& 0.865	\\
\hline
36			& 0.863	\\
\hline
\end{tabular}
\caption{\label{fixed-outcome}Probability $p$ of probe success per parsec necessary to ensure the expected number of systems $\langle N \rangle = 10$ when $N_{target} = 60$ systems are visited.}
\end{table}


\section{Discussion}
\label{conclusions}

In this paper, the characteristics of a future program is looked at for exploring the local interstellar neighborhood with a fixed number of probes, each of which travels at a constant cruise speed and uses only gravitational deflections to change direction. When designing such a program, one crucial metric is the amount of time it takes for the entire program to be completed, or at least the average time for a typical probe to finish its preprogramed trajectory. These values depend on the maximum possible speed of the probes, the number of systems targeted, and the number of probes sent out to do the exploring. Roughly speaking, doubling the number of probes or the speed of those probes serves to half the amount of time to finish the desired exploration, while the completion time is proportional to the target list -- doubling the targets will approximately double the time to visit them. The latter is somewhat surprising, considering that twice the number of targets will fill a volume roughly $2^{1/3} \simeq 1.26$ times as large. These conclusions are codified in the scaling laws (\ref{scaling-mean}) through (\ref{scaling-max}). A second important factor to examine is how the probability of success alters the expected number of targets visited by functioning spacecraft. Somewhat surprisingly, increasing the number of probes launched will increase this expected number only up to a certain amount; further probes will not significantly change the sample of star systems explored. 

This study set out to examine how to optimize an exploration program based on ``large" spacecraft, such as a fleet of Daedalus-class probes. As such, the scaling laws obtained focus on how to maximize the results of such a program. However, it is worth commenting here about considerations for an alternate choice, using a much larger number of ``small" probes, such as a mass-produced collection of solar sails or other beamed-propulsion concepts. It is intuitively likely that if multiple probes are sent out along the same path, with each probe having its own success probability $p$ per parsec, that the expected number of systems reached by these probes will be higher than for a single probe. Stated differently, this means the effective success probability $p_{eff}$ of these multiple probes should be larger than $p$. Although this question has not been extensively studied, preliminary comparisons show that this intuition is true, and that the ratio $(p_{eff} - p) / p$ is in the range $0.01 - 0.1$, depending on the number of extra probes sent along the same trajectory.

Next, future avenues of expanding this model are mentioned. One further direction to consider is the inclusion of the motion of the stars relative to the Solar System. Looking only at radial velocities -- i.e. the component of a star's relative motion towards or away from the Solar System -- many of these are on the order of $10^{-4} c$, within the range of the probes in our mathematical model. Thus, over the course of the simulated interstellar exploration programs, there can be significant changes in both the distances between stars, and the relative order of the distances between systems~\cite{Mat94}. This has two consequences for the scenario in this paper. The first is that the relative distance between star systems may change between the time the probes are launched, and when a particular system is reached by one of the spacecraft. This adds a wrinkle not taken into account in the algorithm presented here, namely the possibility it may be advantageous to change the order that star systems are visited in order to use a favorable alignment. Note that this does not specifically refer to a decreasing distance between two targets, but also whether there is a decrease in the necessary angular deflection for a transit between three targets in sequence, allowing the probe to have a larger cruise speed for its entire journey. The second result of including relative motions is the possibility of increasing the cruise speed of a probe by a fortuitous gravitational assist at a target star~\cite{For-Pap-Kit12}, although a decrease may happen as well. Both of these factors would serve to increase the needed complexity of the program design algorithm used in the simulations, since the cruise speed of all probes and the positions of the target systems would be dynamic variables in the search for optimal paths. As detailed in Appendix \ref{SA}, the current simulated annealing algorithm looks for good paths by sampling random choices, evaluating the cruise speed and time for completion after the choice is made. However, if the target systems move in the simulation, then the cruise speed would depend on the positions of the stars because of the necessary angular deflections, but the positions of a given star system when a probe reaches it can only be calculated when the time the probe arrives there is known, which depends on the probe's cruise speed. The result is that the cruise speed for a given path would have to be optimized, leading to extra calculations in the algorithm.

Finally, the implications are mentioned for the possibility of probes from extraterrestrial civilizations to visit the Solar System. Because the results obtained here focus on optimizing the exploration of a certain volume of interstellar space, there are many instances where probes move back towards the Solar System as they complete their mission. Thus, the progress of the probes does not match a wave front always propagating away from their origin (such as that modeled by Newman and Sagan~\cite{New-Sag85}), so it is somewhat difficult to extrapolate these values beyond the scenarios considered. This can be seen in Figure \ref{cum-time-graph}, showing the cumulative time to complete the program as a function of total number of targets $N_{target}$. Although these times are roughly linear for most of the $N_{target}$ values, there is a marked change around $N_{target} \simeq 55$ for $N_{probe} = 3$ probes. This alteration in behavior may occur for larger numbers of probes, only at greater choices of $N_{target}$. This question can only be decided by explicitly carrying out the required simulations. However, it is viable to say that the exploration program of a nearby civilization would reach the Solar System within a few tens of kiloyears. Another issue is that, for all simulations presented here, it is assumed that all systems within a certain distance threshold are targeted by probes, but this is certainly not the only choice. In particular, if a civilization is only interested in stars of a similar spectral class to their own, there is the chance that their interstellar exploration may not pick our Sun as a target. However, the Solar System may still be visited by an alien probe, solely for the possibility of a favorable angular deflection provided by the Sun between two target systems. Indeed, examining exact solutions using a branch-and-bound method, for 11 total systems and a selection of $N_{target} < 11$, many of the obtained solutions include multiple systems not on the target list~\cite{Car11}, included only for navigational purposes. A comparison of the number of solutions with extra systems to the total number of target lists possible is given in Table \ref{extra-target}. Using the relation (\ref{min-dist-eqn}) for the perihelion $r_p$ of the probe as a function of maximum desired heat flux $H_{max}$, we find that
\begin{equation}
	r_p = (0.037 \ \textrm{AU}) \sqrt{\biggl( \frac{L_s}{L_{\astrosun}} \biggr) \biggl( \frac{1 \ \textrm{MW/m}^2}{H_{max}} \biggr)}
\end{equation}
Thus, for an extraterrestrial probe using the Sun for gravitational deflection, $r_p = 0.014$ AU, well within the orbit of Mercury, and thus easily observed from Earth. This suggests the question of whether there are favorable routes for probes traveling through our Solar System using only gravitational deflections -- in other words, if probes of this nature moving through the Solar System in between nearby stars might be more likely to travel along a small number of trajectories. Computing these paths, and targeting the resulting sectors of the celestial sphere for SETI observations, makes for an interesting future project.

\begin{table}
\begin{tabular}{|c|c|c|}
\hline
$N_{target}$	& Solutions with additional systems	& Total number of target list choices	\\
\hline
\hline
4		& 66					& 330 					\\
\hline
5		& 109					& 462					\\
\hline
6		& 107					& 462					\\
\hline
7		& 24					& 330					\\
\hline
8		& 24					& 165					\\
\hline
\end{tabular}
\caption{\label{extra-target}Number of exact solutions featuring systems not on the target list, compared to the total number of choices for the target list, for $N_{target}$ systems out of a possible 11 targets~\cite{Car11}.}
\end{table}

\section{Acknowledgements}

The author wishes to thank Duncan Forgan for helpful comments on a draft version of this work.

\appendix

\section{Simulation algorithms}
\label{SA}

The algorithm used to compute the simulations presented in this paper is detailed here, along with a discussion of simpler heuristic methods, and a comparison of these methods to exact solutions. All the numerical results shown use the standard numerical technique of simulated annealing in order to arrive at close to optimal solutions for large data sets in a reasonable amount of time.  Obtaining exactly optimal solutions requires more computational effort, for example, by a branch-and-bound search through likely candidates. However, using simpler methods than simulated annealing may allow for decent solutions obtained in even faster time; note that many of the previous efforts~\cite{bjork, Cot-Mor09, For-Pap-Kit12} use these heuristic methods to obtain results, although Cotta and Morales supplement these with various improvement techniques. Below, the simulated annealing algorithm used in this paper is detailed, as well as two heuristic methods, and compare these results to those of exact solution obtained by a branch-and-bound method. The interested reader may turn to Dasgupta, Papadimitriou, and Vazirani~\cite{algo-book} for further information about these algorithmic methods.

Simulated annealing is a technique inspired by the physical process of controlling heating and cooling of a material in order to form large-sized crystals~\cite{sim-anneal}. The process starts with an initially random solution, and its fitness -- i.e. the value of the given objective function -- is evaluated. This is compared to the fitness of ``nearby" solutions, where the initial solution is changed slightly by a random process. If the new solution is better, it is kept in lieu of the original solution.  This is analogous to the heated material, where atoms may shift around at large temperatures and improve the crystalline structure, removing defects. However, it is possible that the shift may increase the energy of the material system, for increases less than a temperature-dependent function. The higher the temperature, the more these jumps are allowed. In a like manner, the simulated annealing algorithm may keep marginally worse solutions, in order to test out the solution space for deeper minima of the objective function.
As the temperature decreases, these possible shifts are not as large, as it is hoped the material has reached a local minimum in its free energy. Similarly, for the simulated annealing process, the amount of change allowed to the current solution decreases, based on a decreasing parameter analogous to temperature. In practice, multiple trials of the algorithm are necessary to ensure that a reasonable swath of the solution space is searched effectively.

The particular version of the algorithm used here is detailed in pseudocode in Algorithm \ref{algo}. In order to find the best path, the simulated annealing algorithm is run for a number of trials, typically $Trials = 60$. The objective function $C$ used is the Euclidean norm $\sum_a t^2 _a$, with $t_a$ the time it takes for probe $a$ to complete its particular mission. For $n = N_{target}$ target systems, each path is stored as an array giving a permutation of the integers in the range $[1, n]$ -- the order the probe visits each system $i$ after it leaves the source node $i = 0$. When there is more than one probe $N_{probe} = P$ considered, these additional probes are assigned numbers in the range $[n + 1, n + P - 1]$ and the path array is a permutation of $[1, n + P - 1]$; the upper bound is $n + P - 1$ since the source node $i = 0$ also counts as a ``probe" node. The extra nodes are taken to have the same position as the source node for purposes of calculating the objective function of the overall probe collection. Since all probes are ignored after they visit the last target system on their list, there is no contribution to the objective function for portions of the path array going from a target system to a ``probe" node. A trial starts with a random permutation of these integers; nearby solutions are obtained with $ListSwitch$ by switching two of the numbers in the path array, and stored in the array $tempList$. The results of each individual trial are stored in the temporary list $tempMinPath$, and compared to the overall minimum norm path $minPath$; if $tempMinPath$ has a lower objective function, it is kept as the new best solution. In this implementation, the temperature schedule used is $1 / \log(k + 1)$, for step $k$ of the algorithm, up to a maximum of $(n + 1)^2$ steps. This choice gives reasonably good results in a short period of time.

As stated before, the simulated annealing algorithm is not guaranteed to calculate the optimal solution, but is a method to quickly arrive at one that is close to optimal. To find the exact optimal solution, one must use other techniques such as brute-force search or a branch-and-bound method; the use of the latter in the interstellar exploration program will be detailed in a future paper. On the other hand, one can use heuristic techniques which arrive at potential solutions even more quickly than simulated annealing, at the cost of larger differences with the cost of the solution compared to the optimal one. Below are detailed two possible heuristic greedy algorithms each probe can use to decide which star system to visit next; as with the simulated annealing algorithm, they pre-calculate the path, and require the knowledge of coordinates and stellar characteristics of all possible targets.
\begin{enumerate}

	\item {\it Fastest speed:} For each leg of the journey, coming from system $i$ to system $j$, the next system $k$ after $j$ is chosen for the largest possible cruise speed out of all possibilities $i \to j \to k$. This depends only on the deflection angle for the trajectory $i \to j \to k$.
	
	\item {\it Shortest time:} As the probe arrives at system $j$ after the leg $i \to j$, the next system $k$ is chosen so that the leg $j \to k$ has the shortest possible travel time of all possible targets. Note this is a function both of the deflection angle for the path $i \to j \to k$ and the distance from $j$ to $k$.

\end{enumerate}
In both cases, as the path is assembled from each portion, the allowed cruise speed may be reduced, because the required angular deflection at a given system may decrease the speed upper bound. Because these two algorithms do not test that many choices, compared to the simulated annealing algorithm, they are both much faster to compute. However, there is a corresponding decrease in optimality. To show this explicitly, the simple case of $N_{probe} = 1$ and small numbers of target stars ($N_{target} \le 14$) is considered. In Table \ref{exact-SA-comparison}, the simulated annealing and heuristic algorithms detailed above are compared to the exact results, and the percent excess in the time for the single probe to complete its mission are given. One can see that both fare rather poorly compared to the optimal solution, and even the simulated annealing algorithm, which works rather well for these test cases.

\begin{table}[hbt]
\begin{tabular}{|c|c|c|c|}
\hline
$N_{target}$	& Simulated annealing	& Fastest speed	& Shortest time\\
\hline
\hline
9			& 5.30 \%				& 80.2 \%			& 3.93 \%		\\
\hline
10			& 24.6 \%				& 123 \%			& 18.5 \%		\\
\hline
11			& 14.2 \%				& 17.7 \%			& 117 \%		\\
\hline
12			& 7.29 \%				& 130 \%			& 105 \%		\\
\hline
13			& 17.8 \%				& 145 \%			& 45.7 \%		\\
\hline
14			& 18.0 \%				& 72.4 \%			& 53.0 \%		\\
\hline
\end{tabular}
\caption{\label{exact-SA-comparison}Estimation errors in the time for a single probe ($N_{probe} = 1$) to complete its mission for the simulated annealing algorithm, and two heuristic greedy algorithms -- fastest speed and shortest time -- compared to the exact optimal solutions obtained using a branch-and-bound algorithm, with $N_{target}$ given in the table. For the simulated annealing algorithm, a total of 60 trials were used. When there are eight or less target star systems, the exact and simulated annealing solutions agree. Each of the three algorithms is as described in the Appendix.}
\end{table}

\begin{pseudocode}[ruled]{Simulated Annealing}{{\vec r}_i}
\label{algo}
	minTime \GETS \infty										\\
	minPath \GETS \{\}											\\
	\\
	\FOR i \GETS 1 \TO Trials \DO
		\BEGIN
			probeTrail \GETS \CALL{MakeRandomTrail}{n + P}			\\
			E_{old} \GETS \CALL{ObjectiveFunction}{probeTrail}		\\
			k_{max} \GETS (n + 1)^2								\\
			\\
			tempMinTime \GETS \infty								\\
			tempMinPath \GETS \{\}								\\
			\\
			\FOR k \GETS 1 \TO k_{max} \DO
				\BEGIN
					tempList \GETS \CALL{ListSwitch}{probeTrail}		\\
					E_{new} \GETS \CALL{ObjectiveFunction}{probeTrail} \\
					T \GETS 1 / \log(k + 1)						\\
					\\
					\IF \exp[(E_{old} - E_{new}) / T] > \CALL{Random}{}	\THEN
						\BEGIN
							probeTrail \GETS tempList			\\
							E_{old} \GETS E_{new}				\\
						\END
					\\
					\IF E_{new} < tempMinTime \THEN
						\BEGIN
							tempMinTime \GETS E_{new}			\\
							tempMinPath \GETS tempList			\\
						\END
				\END
				\\
				\IF tempMinTime < minTime \THEN
					\BEGIN
						minTime \GETS tempMinTime				\\
						minPath \GETS tempMinPath				\\
					\END
		\END
	\\
	\\
	\OUTPUT{minPath}
\end{pseudocode}

\end{document}